\documentclass{article}
\usepackage{spconf,amsmath,graphicx,hyperref}
\usepackage{amssymb}
\usepackage{multirow}

\hypersetup{
    colorlinks=true,
    citecolor=blue,
    linkcolor=blue,
    filecolor=magenta,      
    urlcolor=blue,
}
\title{Deep model with built-in cross-attention alignment for acoustic echo cancellation}
\name{Evgenii Indenbom, Nicolae-C\u{a}t\u{a}lin Ristea, Ando Saabas, Tanel P\"arnamaa, Jegor Gužvin, Ross Cutler}

\address{Microsoft Corporation}

\begin{document}
%
\maketitle
\begin{abstract}
With recent research advances, deep learning models have become an attractive choice for acoustic echo cancellation (AEC) in real-time teleconferencing applications. Since acoustic echo is one of the major sources of poor audio quality, a wide variety of deep models have been proposed. However, an important but often omitted requirement for good echo cancellation quality is the synchronization of the microphone and far end signals. Typically implemented using classical algorithms based on cross-correlation, the alignment module is a separate functional block with known design limitations. In our work, we propose a deep learning architecture with built-in cross-attention-based alignment, which is able to handle unaligned inputs, improving echo cancellation performance while simplifying the communication pipeline. We show that our approach achieves significant improvements for difficult delay estimation cases on real recordings from the AEC Challenge dataset. Moreover, our model is state-of-the-art for low-complexity AEC, and has been successfully deployed in Microsoft Teams for millions of users.
\end{abstract}
\vspace{-0.2cm}
\section{Introduction}
\vspace{-0.2cm}

In recent years, the use of teleconferencing systems, e.g. Microsoft Teams, Skype, Zoom, etc., has increased significantly, becoming indispensable for remote work. For these systems, it is mandatory to ensure good call quality in order to provide a productive and pleasant experience to end-users. Degradations caused by acoustic echo are one of the major sources of poor call quality in audio and video calls. This problem becomes even more challenging in full duplex communication when echo interferes with double talk scenarios \cite{sridhar2021icassp}. Formally, acoustic echo occurs when a loudspeaker and a microphone are coupled in a communication system such that the microphone picks up the loudspeaker signal, usually distorted by the room impulse response. If not properly handled, the far end user will hear their own voice delayed by the round trip time of the system (i.e., an echo), mixed with the signal from the near end. To solve this, AEC algorithms are employed. 


The application of deep learning models for audio-related tasks started with combining classical digital signal processing (DSP) based methods with neural networks. For example, \cite{zhang2019deep, ma2020acoustic, zhang2021deep, zhang2022deep} have demonstrated that the combination of adaptive filters and recurrent neural networks (RNNs) provides good performance in the AEC task. Other researchers have tried to develop a pure deep learning solution for the echo cancellation problem and obtained convincing results \cite{sridhar2021icassp, fazel2020cad, cutler2021interspeech, zhao2022deep} on complex datasets. The ability of RNNs (e.g., gated recurrent units (GRU) \cite{chung2014empirical}, long short term memory (LSTM) \cite{hochreiter1997long}) to model time-varying functions plays an important role in addressing AEC problems \cite{braun2021towards, westhausen2021acoustic, zhao2022deep}. Braun et al.~\cite{braun2021towards} proposed the CRUSE model for noise suppression, based on a U-Net architecture, with a middle recurrent block. The model offers a good trade-off between computational complexity and speech quality, measured on real recordings using a mean opinion score (MOS). Motivated by these advances in applying deep learning models in audio tasks, we seek to provide a model which can simplify the communication pipeline, while improving the overall performance.

\begin{figure}[!t]
\begin{center}
\centerline{\includegraphics[width=0.8\linewidth]{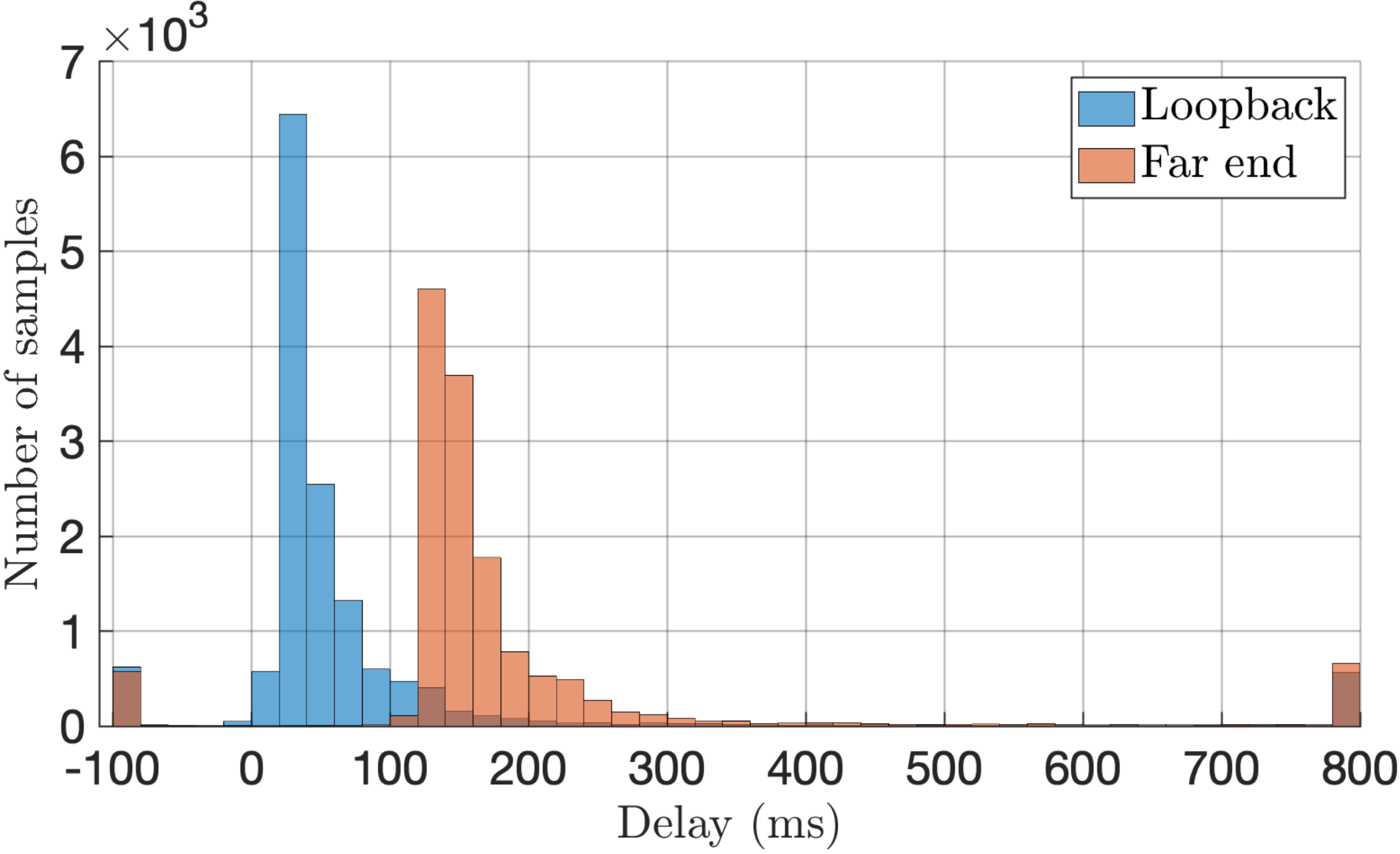}}
\caption{Delay distribution for far end and loopback signals compared to the microphone. Delays smaller than $-0.1$ms and bigger than $0.8$ms are truncated.}
\label{fig_echo}
\end{center}
\vskip -1.2cm
\end{figure}

One major issue in AEC systems is the delay between the microphone and the far end reference signals, which drastically affects performance. In most cases, it is assumed that the acoustic echo path is linear and the time delay is limited to a known prior. Under this assumption, the acoustic echo signal can be canceled effectively using traditional methods \cite{farhang2013adaptive}. However, the performance of existing linear AEC algorithms may be greatly degraded in many practical applications, because of the hardware-related latency or software buffering mechanism, which leads to a larger range of delays. Moreover, scenarios when the acoustic echo path may be time varying (e.g., the speaker changes location during a call), or the microphone signal could contain some environmental noise and reverberation also impact the AEC performance.
Fig.~\ref{fig_echo} shows the distribution of estimated delays between the microphone and both far end and loopback signals on real recordings from the AEC Challenge dataset \cite{cutler2021interspeech}. The delays are estimated using a cross-correlation-based algorithm over the entire clip. The plot shows a long tail for the delay distributions, but also outliers in both ends of the distributions, i.e., estimates of negative delay or delay over $800$ ms. The outliers are typically misestimated because of distorted or noisy signals. In real-time call scenarios, such misestimates typically result in echo leaks when conventional algorithms for signal alignment are used.

Many works use already aligned data or compensate for the delay by a separate block \cite{peng2021icassp, watcharasupat2021end}, a procedure that can work poorly in practice. A different approach is proposed in \cite{ma2021echofilter}, where the authors developed a model which performs alignment in the time domain with a local attention block. The attention block computes the alignment based on the RNN's internal states. Distinctly, our model performs attention on deep time-frequency features, computing an actual delay distribution, which is used to soft align the far end features.


In this work, we propose a real-time deep neural network architecture with a built-in alignment module based on cross-attention, named Align-CRUSE, capable of handling non-aligned microphone and far end signals in linear and non-linear echo path scenarios. Our model eliminates the necessity of an external alignment block, which is conventionally based on DSP algorithms (e.g., cross-correlation \cite{ianniello1982time}), by including a built-in cross-attention module that synchronizes the microphone and far end signals in the feature space. 
Instead of the alignment of signals, the module performs \emph{soft} alignment via attention, where multiple delay estimates can be used in difficult delay scenarios.
The use of soft in-model alignment improves the AEC quality and archives good results in complex real-world scenarios. Furthermore, considering the low computational complexity and the low inference time and latency, our model was successfully deployed in Microsoft Teams, obtaining significant improvements for millions of calls every day.

\vspace{-0.2cm}
\section{Proposed Method}
\vspace{-0.2cm}

\subsection{Problem formulation}
\vspace{-0.2cm}

A generic AEC system could be formally described as follows. The far end reference signal is transmitted to the receiving room, played back through the loudspeaker, and then picked up by the microphone signal via an acoustic echo path (modeled by a room impulse response). The captured microphone signal, composed of a near end signal, background noise, and echo is received by the far end user.
We highlight that the echo component from the microphone signal is a delayed version of the received reference far end signal, because of the echo propagation path (from loudspeaker to microphone) and hardware or software-related latency. Therefore, between the far end and near end users an AEC system is integrated. Hence, our goal is to remove undesired echoes, having the microphone and the reference far end signals.



\vspace{-0.35cm}
\subsection{Feature extraction}
\vspace{-0.2cm}

All audio signals are sampled at 16 kHz and the preprocessing is performed identically for the reference far end and microphone signals. The input features to the network are log power spectra computed with a squared root Hann window.

\vspace{-0.35cm}
\subsection{Network architecture}
\vspace{-0.2cm}

\begin{figure*}[!t]
\begin{center}
\centerline{\includegraphics[width=0.7\linewidth]{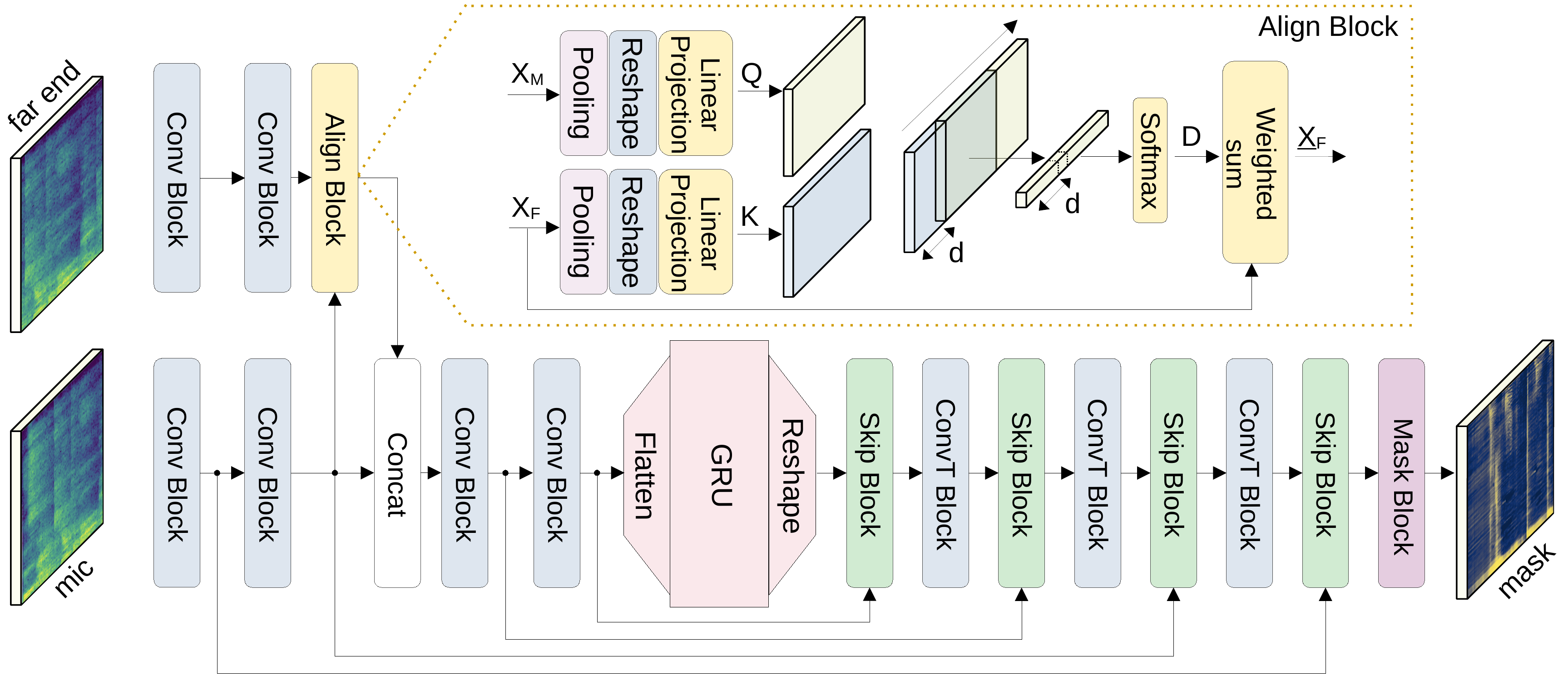}}
\caption{Align-CRUSE architecture for the AEC task. The output is a magnitude mask, used to predict the enhanced microphone signal. In the top, the align block is illustrated. Best viewed in colors.}
\label{fig_alignCruse}
\end{center}
\vskip -1.0cm
\end{figure*}

The network structure is derived from \cite{braun2021towards}, with AEC adaptation and a cross-attention mechanism for built-in deep alignment. The overall model architecture is illustrated in Fig.~\ref{fig_alignCruse}. In this section we will use $c, t, f \in \mathbb{N}$ to denote arbitrary channel, time and frequency axis lengths.



\noindent \textbf{Encoder.} 
The encoding part consists of two branches, having as input the far end and microphone features. Each conv block is built by a convolutional layer, a batch-norm layer, and an exponential linear unit (ELU) \cite{clevert2015fast} activation function. The far end branch is composed of two conv blocks, followed by the align block, which takes as input the far end feature maps and the depth-corresponding feature maps from the mic branch. Consequently, the aligned far end maps are concatenated to the mic branch and processed by two more conv blocks. The numbers of kernels for the microphone conv blocks are 16, 40, 72, and 32. The far end branch is composed of 8 and 24 filters. All convolution kernels have size $4 \times 3$ and a stride of $1 \times 2$, reducing the number of bins along the frequency axis. Each convolution is causal, meaning that the padding is performed such that no look-ahead is used.

\noindent \textbf{Recurrent block.}
Between the encoder and decoder sits a recurrent layer, which is fed with feature maps flattened along the channel and frequency dimensions. Formally, the input $X\in \mathbb{R}^{c \times t \times f}$ is flattened into $X \in \mathbb{R}^{t \times (cf)}$. Afterward, $X$ is fed into the recurrent layer and the output is reshaped back to $X\in \mathbb{R}^{c \times t \times f}$. Following \cite{braun2021towards}, considering that an LSTM does not bring significant performance improvements, we use a GRU layer to reduce the model complexity. 


\noindent \textbf{Skip block.}
Replacing the classical skip connection, based on concatenation or summing, with a trainable channel-wise scaling and bias, improves the performance at a small additional cost \cite{braun2021towards}. Moreover, it allows us to use asymmetric encoder-decoder blocks by adapting the number of encoding channels to the number of corresponding decoding channels. Therefore, each skip block is composed of a convolutional layer with the same number of kernels as the depth of the corresponding decoder input and a kernel size of $1\times 1$ applied with a stride of $1 \times 1$.

\noindent \textbf{Decoder.} 
The decoding stage consists of $3$ alternating skip and conv transpose blocks, followed by a last skip and mask block. The reshaped output from the GRU is combined with the corresponding features from the encoder (into the skip block) and fed into the transpose conv block. Each transpose conv block is composed of a transposed convolutional layer, followed by a batch-norm and an ELU activation function. For each transpose convolutional layer, we used a causal kernel, with a size of $1 \times 3$, which computes features along the frequency dimension. The stride is identical to the encoding part, while the number of filters for convolutional blocks is 32, 48, and 48. Subsequently, the output of the last skip block is processed by the mask block. It consists of a convolutional layer with a single filter of $1 \times 3$, followed by a sigmoid activation. Moreover, to compensate for potential over-suppression by applying magnitude masking, we added a learnable parameter in the mask block, which multiplies with the output mask. This allows the model to have an internal learnable gain control.

\noindent \textbf{Align block.} 
Let $X_M\in \mathbb{R}^{c \times t \times f}$ be the mic features and $X_F\in \mathbb{R}^{c \times t \times f}$ the far end features. Firstly, the feature maps are reduced with a max-pooling layer (having a kernel size of $1 \times 4$) along the frequency dimension to reduce the computation cost brought by the alignment module. Further, the features are reshaped such that $X_M'\in \mathbb{R}^{t \times (\frac{f}{4} \cdot c)}$ and $X_F'\in \mathbb{R}^{t \times (\frac{f}{4} \cdot c)}$. Next, they are projected into queries $Q \in \mathbb{R}^{t \times p}$ and keys $K \in \mathbb{R}^{t \times p}$, where $p \in \mathbb{N}$ is the projection size. Secondly, the $K$ tensor is zero-padded at the beginning and cropped at the end with the same $d$ value, generating a synthetic delay. Afterward, a time axis dot product is performed for  $Q$ and delayed $K$. This procedure is computed for each delay index $d$ from a specific interval, given by the maximum supported delay $d_{max}$, leading to a result vector of length $d_{max}$. The vector is further used in softmax activation, predicting the most likely delay distribution $D \in \mathbb{R}^{d_{max}}$. Afterward, the aligned far end features $\underline{X}_F\in \mathbb{R}^{c \times t \times f}$ are computed through a weighted sum on the time axis with the corresponding delay probability from $D$. More precisely, $X_F$ is delayed, multiplied with the corresponding weight factor from $D$, and added to the final result  $\underline{X}_F$.
We observed that having a weighted sum rather than a hard selection improves the robustness to wrong delay estimates by allowing flat delay distributions. Our novel aligning block is presented in the upper part of  Fig.~\ref{fig_alignCruse}.





\begin{table*}[!t]
\caption{Results for Align-CRUSE model on LD-M, LD-H, and AEC challenge \cite{cutler2021interspeech} single talk test set, split into FEST-HD and FEST-GEN. The baseline CRUSE model was tested on non-aligned data (CRUSE), online aligned data (CRUSE$^\ddagger$), and globally aligned data (CRUSE$^{\ddagger \ddagger}$). 
We also reported the $95\%$ confidence interval. The inference time is in milliseconds per frame.}
\label{tab_results_align}
\setlength\tabcolsep{1.0pt}
\centering
\begin{tabular}{l | c c | c c | c c | c c | c c}
{Method} & \multicolumn{2}{|c|}{\multirow{2}{0pt}{}{{LD-M}}} &
\multicolumn{2}{|c|}{\multirow{2}{0pt}{}{{LD-H}}} &
\multicolumn{2}{|c}{\multirow{2}{0pt}{}{{FEST-HD \cite{cutler2021interspeech}}}} &
\multicolumn{2}{|c}{\multirow{2}{0pt}{}{{FEST-GEN \cite{cutler2021interspeech}}}} &
\multicolumn{2}{|c}{\multirow{2}{0pt}{}{{Inference}}} \\

& {AECMOS$\uparrow$} & {ERLE$\uparrow$} &  {AECMOS$\uparrow$} & {ERLE$\uparrow$}  &  {AECMOS$\uparrow$} & {ERLE$\uparrow$} &  {AECMOS$\uparrow$} & {ERLE$\uparrow$} & {Time(ms)} & {\#Params}\\  
\hline

CRUSE                   & 2.31\scriptsize{$\pm 0.06$} & 7.63\scriptsize{$\pm 0.62$} & 2.21\scriptsize{$\pm 0.05$} & 3.93\scriptsize{$\pm 0.39$} & 2.91\scriptsize{$\pm 0.32$} & 19.80\scriptsize{$\pm 4.81$} & 4.32\scriptsize{$\pm 0.08$} & 42.43\scriptsize{$\pm 1.96$} & 0.216 & 0.74M\\
CRUSE $^\ddagger$       & 3.89\scriptsize{$\pm 0.06$} & 26.30\scriptsize{$\pm 1.26$} & 3.57\scriptsize{$\pm 0.06$} & 17.57\scriptsize{$\pm 0.98$} & 3.78\scriptsize{$\pm 0.32$} & 34.54\scriptsize{$\pm 4.81$} & 4.42\scriptsize{$\pm 0.08$} & 42.55\scriptsize{$\pm 1.96$} & 0.216& 0.74M\\
CRUSE  $^{\ddagger \ddagger}$                  & 3.99\scriptsize{$\pm 0.09$} & 35.84\scriptsize{$\pm 1.54$} & 3.84\scriptsize{$\pm 0.10$} & 33.50\scriptsize{$\pm 1.74$} & 4.30\scriptsize{$\pm 0.18$} & 45.58\scriptsize{$\pm 3.85$} & 4.38\scriptsize{$\pm 0.05$} & 43.38\scriptsize{$\pm 1.71$} & 0.216 & 0.74M\\
\hline
Align-CRUSE             & \textbf{4.54}\scriptsize{$\pm 0.01$} & \textbf{42.88}\scriptsize{$\pm 1.15$} & \textbf{4.44}\scriptsize{$\pm 0.03$} & \textbf{39.37}\scriptsize{$\pm 1.28$} & \textbf{4.54}\scriptsize{$\pm 0.14$} & \textbf{52.80}\scriptsize{$\pm 4.09$} & \textbf{4.46}\scriptsize{$\pm 0.06$} & \textbf{45.82}\scriptsize{$\pm 1.94$} & 0.218 &0.75M\\


\hline
\end{tabular}
\vspace{-0.3cm}
\end{table*}



\begin{table}[!t]
\caption{Results for Align-CRUSE model on the AEC \cite{cutler2021interspeech} test set, against the challenge's winner and the CRUSE$^\ddagger$. We reported MOS with $95\%$ confidence interval for DT and FEST.}
\label{tab_results_mos}
\centering
\begin{tabular}{l | c c c}
{Method} & {\multirow{2}{0pt}{}{{FEST}}} & {\multirow{2}{0pt}{}{{DT Echo}}} & {\multirow{2}{0pt}{}{{DT Other}}} \\ 
& {MOS$\uparrow$} & {MOS$\uparrow$} & {MOS$\uparrow$} \\

\hline
AEC winner \cite{peng2021acoustic}                    & 4.34\scriptsize{$\pm 0.02$} & 4.36\scriptsize{$\pm 0.02$} & \textbf{4.23}\scriptsize{$\pm 0.02$}\\
\hline
CRUSE$^\ddagger$                     & 4.55\scriptsize{$\pm 0.02$} & 4.42\scriptsize{$\pm 0.02$} & 4.07\scriptsize{$\pm 0.02$}\\
Align-CRUSE                        & \textbf{4.67}\scriptsize{$\pm 0.02$} & \textbf{4.45}\scriptsize{$\pm 0.02$} & 4.07\scriptsize{$\pm 0.02$} \\

\hline
\end{tabular}
\vspace{-0.4cm}
\end{table}

\noindent \textbf{Prediction.} 
The model's output is a suppression magnitude mask used to enhance the mic signal by removing undesired components. The inference process starts with the compressed time-frequency features of the mic and far end signals. The output suppression mask is multiplied by the microphone's magnitude spectrum (preserving the original phase), resulting in the enhanced spectrum. Further, the inverse short-time Fourier transform (STFT) is computed to obtain the enhanced time domain signal.

\vspace{-0.35cm}
\subsection{Loss function}
\vspace{-0.2cm}
We train the networks with STFT consistency enforcement \cite{wisdom2019differentiable} by propagating the time domain enhanced output again through STFT. The network is optimized by minimizing the complex compressed mean-squared error loss \cite{ephrat2018looking}, which blends the magnitude with a phase-aware term, which we found to be superior to other losses. We used the compression factor $c = 0.3$ and the weighting factor between complex and magnitude-based losses $\beta = 0.7$.

\vspace{-0.35cm}
\section{Experimental setup}
\vspace{-0.2cm}


\noindent \textbf{Datasets.}
To ensure the generalization ability, the training data are synthesized online with random parameters for each sample (e.g., signal-to-noise ratio, distortion, gain, signal-to-echo ratio). We trained the networks on training data from the AEC Challenge \cite{cutler2021interspeech}. 
We tested our approach on the blind test set from the AEC Challenge, which contains real-world recordings in diverse scenarios. In Table \ref{tab_results_align}, we split the far end single talk (FEST) blind test set into two: FEST-HD, which contains 27 samples with difficult delay estimation cases (e.g., long or variable delays, as indicated by the authors), and FEST-GEN, which contains 273 samples with other types of scenarios (e.g., non-linear-distortions, stationary-noise). In addition, we generated two synthetic datasets, containing 500 samples each, to specifically address long delay cases. In LD-M we randomly distributed the delays uniformly in $[0.3,0.5]s$, while for LD-H in $[0.5,1.0]s$.



\noindent \textbf{Evaluation metrics.}
We employed echo AECMOS \cite{purin2021aecmos} to test the removal capacity of echos. 
For echo cancellation ability in FEST scenarios, we also employed the echo return loss enhancement (ERLE). Moreover, we submitted the enhanced samples to human raters and obtained MOS scores \cite{cutler2021crowdsourcing}. Each sample was annotated by 5 distinct raters.


\noindent \textbf{Hyper-parameters tuning.}
For feature generation, we used a squared root Hann window of length $20$ms, a hop length of $10$ms and a discrete Fourier transform length of 320. This leads to an overall algorithmic latency of 20ms. We trained both CRUSE and Align-CRUSE models in the same fashion. For both models, we used the Adam \cite{kingma2015adam} optimizer, with batches of $400$ samples for $150$ epochs, with a learning rate of $1.5 \cdot 10^{-4}$ and a weight decay of $5 \cdot 10^{-6}$. We considered $d_{max}=100$, equivalent to a maximum delay of $1$ second.

\noindent \textbf{Baseline.}
We compare the Align-CRUSE against CRUSE adapted for the AEC task by having two input channels, mic and previously aligned far end features. We used the baseline on unaligned (CRUSE), real-time aligned (CRUSE$^\ddagger$), and globally aligned (CRUSE$^{\ddagger\ddagger}$) far end data. Both real-time and global alignment methods are based on cross-correlation. For the real-time method \cite{ianniello1982time}, the alignment is done frame by frame, considering past data frames (no look-ahead), while for the global method the delay is estimated considering the entire audio. We included the last method as a stronger baseline, though it is unsuitable for real-time application. 
Moreover, we included the AEC Challenge winner \cite{peng2021acoustic}.

\noindent \textbf{Results.}
On synthetic datasets, our model surpasses CRUSE$^{\ddagger\ddagger}$ by up to $0.6$ AECMOS and $7$ dB ERLE. Against other baseline models, the difference is even more significant.
To test the generalization capacity of our model, we applied it on real-world recordings from FEST-HD and FEST-GEN datasets. Compared to CRUSE$^\ddagger$, the improvement is $7.22$ dB ERLE and $0.24$ AECMOS for FEST-HD and $2.44$ dB ERLE and $0.08$ AECMOS for FEST-GEN.  Overall, the results show that having a robust delay estimator is mandatory for hard delay estimation cases, a fact also observed in production. Moreover, we included comparative audio samples and visualizations of the alignment feature map in \url{https://ristea.github.io/aec-align-cruse}.


In addition to objective metrics, we compared the models on the AEC Challenge \cite{cutler2021interspeech} blind set in terms of double-talk (DT) and FEST MOS scores (Table \ref{tab_results_mos}). For this test, we included the challenge's winner \cite{peng2021acoustic} and only CRUSE$^\ddagger$ as a baseline, since CRUSE tested on unaligned data is considerably worse (see Table \ref{tab_results_align}) and CRUSE$^{\ddagger\ddagger}$ is not feasible for real-time applications. We observe significant improvements for both FEST MOS and DT Echo MOS. Even if \cite{peng2021acoustic} attains better results in terms of DT Other MOS, the model is around $5$-$7$ times slower compared to ours and has an algorithmic latency of 30ms instead of 20ms.



Additionally, Table \ref{tab_results_align} shows the number of parameters and the inference time per frame on a CPU Intel Core i7 11370H@3.3 GHz. Our model has roughly the same inference time as the CRUSE model and only $0.01$M parameters more. Thus, Align-CRUSE provides significantly better results, especially in long and hard delay estimation cases, while assuring about the same inference time, a critical aspect for real-time processing. We included extended performance comparisons on the project's web page. 


\vspace{-0.5cm}

\section{Conclusion}
\vspace{-0.3cm}
We proposed the Align-CRUSE architecture with a built-in aligning module, based on cross-attention, which is able to handle unaligned far end and mic signals in complex echo scenarios. We empirically demonstrated the alignment importance in AEC systems for both synthetic and real-world audio samples. Align-CRUSE improves the AEC in both single talk and DT scenarios while simplifying the communication pipeline, concluding that it is the state-of-the-art method in low-complexity real-time AEC. Moreover, the model has been successfully deployed in MS Teams, enhancing the quality for millions of calls every day. In future work, we aim to develop the model to better address DT scenarios.

\bibliographystyle{IEEEtran}

\bibliography{mybib}

\begin{thebibliography}{10}
\providecommand{\url}[1]{#1}
\csname url@samestyle\endcsname
\providecommand{\newblock}{\relax}
\providecommand{\bibinfo}[2]{#2}
\providecommand{\BIBentrySTDinterwordspacing}{\spaceskip=0pt\relax}
\providecommand{\BIBentryALTinterwordstretchfactor}{4}
\providecommand{\BIBentryALTinterwordspacing}{\spaceskip=\fontdimen2\font plus
\BIBentryALTinterwordstretchfactor\fontdimen3\font minus
  \fontdimen4\font\relax}
\providecommand{\BIBforeignlanguage}[2]{{%
\expandafter\ifx\csname l@#1\endcsname\relax
\typeout{** WARNING: IEEEtran.bst: No hyphenation pattern has been}%
\typeout{** loaded for the language `#1'. Using the pattern for}%
\typeout{** the default language instead.}%
\else
\language=\csname l@#1\endcsname
\fi
#2}}
\providecommand{\BIBdecl}{\relax}
\BIBdecl

\bibitem{sridhar2021icassp}
K.~Sridhar, R.~Cutler, A.~Saabas, T.~Parnamaa, M.~Loide, H.~Gamper, S.~Braun,
  R.~Aichner, and S.~Srinivasan, ``{ICASSP 2021 Acoustic Echo Cancellation
  Challenge: Datasets, Testing Framework, and Results},'' in \emph{Proceedings
  of ICASSP}.\hskip 1em plus 0.5em minus 0.4em\relax IEEE, 2021, pp. 151--155.

\bibitem{zhang2019deep}
H.~Zhang, K.~Tan, and D.~Wang, ``{Deep Learning for Joint Acoustic Echo and
  Noise Cancellation with Nonlinear Distortions},'' in \emph{Proceedings of
  INTERSPEECH}, 2019, pp. 4255--4259.

\bibitem{ma2020acoustic}
L.~Ma, H.~Huang, P.~Zhao, and T.~Su, ``{Acoustic Echo Cancellation by Combining
  Adaptive Digital Filter and Recurrent Neural Network},'' \emph{Proceedings of
  INTERSPEECH}, 2020.

\bibitem{zhang2021deep}
H.~Zhang and D.~Wang, ``{A Deep Learning Approach to Multi-Channel and
  Multi-Microphone Acoustic Echo Cancellation},'' \emph{Proceedings of
  INTERSPEECH}, pp. 1139--1143, 2021.

\bibitem{zhang2022deep}
H.~Zhang, S.~Kandadai, H.~Rao, M.~Kim, T.~Pruthi, and T.~Kristjansson, ``{Deep
  Adaptive AEC: Hybrid of Deep Learning and Adaptive Acoustic Echo
  Cancellation},'' in \emph{Proceedings of ICASSP}.\hskip 1em plus 0.5em minus
  0.4em\relax IEEE, 2022, pp. 756--760.

\bibitem{fazel2020cad}
A.~Fazel, M.~El-Khamy, and J.~Lee, ``{CAD-AEC: Context-Aware Deep Acoustic Echo
  Cancellation},'' in \emph{Proceedings of ICASSP}.\hskip 1em plus 0.5em minus
  0.4em\relax IEEE, 2020, pp. 6919--6923.

\bibitem{cutler2021interspeech}
R.~Cutler, A.~Saabas, T.~Parnamaa, M.~Loida, S.~Sootla, H.~Gamper
  \emph{et~al.}, ``{INTERSPEECH 2021 Acoustic Echo Cancellation Challenge},''
  in \emph{Proceedings of INTERSPEECH}, 2021.

\bibitem{zhao2022deep}
H.~Zhao, N.~Li, R.~Han, L.~Chen, X.~Zheng, C.~Zhang, L.~Guo, and B.~Yu, ``{A
  Deep Hierarchical Fusion Network for Fullband Acoustic Echo Cancellation},''
  in \emph{Proceedings of ICASSP}.\hskip 1em plus 0.5em minus 0.4em\relax IEEE,
  2022, pp. 9112--9116.

\bibitem{chung2014empirical}
J.~Chung, C.~Gulcehre, K.~Cho, and Y.~Bengio, ``{Empirical Evaluation of Gated
  Recurrent Neural Networks on Sequence Modeling},'' in \emph{Proceedings of
  NIPS}, 2014.

\bibitem{hochreiter1997long}
S.~Hochreiter and J.~Schmidhuber, ``{Long Short-Term Memory},'' \emph{Neural
  Computation}, vol.~9, no.~8, pp. 1735--1780, 1997.

\bibitem{braun2021towards}
S.~Braun, H.~Gamper, C.~K. Reddy, and I.~Tashev, ``{Towards Efficient Models
  for Real-Time Deep Noise Suppression},'' in \emph{Proceedings of
  ICASSP}.\hskip 1em plus 0.5em minus 0.4em\relax IEEE, 2021, pp. 656--660.

\bibitem{westhausen2021acoustic}
N.~L. Westhausen and B.~T. Meyer, ``{Acoustic Echo Cancellation With the
  Dual-Signal Transformation LSTM Network},'' in \emph{Proceedings of
  ICASSP}.\hskip 1em plus 0.5em minus 0.4em\relax IEEE, 2021, pp. 7138--7142.

\bibitem{farhang2013adaptive}
B.~Farhang-Boroujeny, \emph{{Adaptive Filters: Theory and Applications}}.\hskip
  1em plus 0.5em minus 0.4em\relax John Wiley \& Sons, 2013.

\bibitem{peng2021icassp}
R.~Peng, L.~Cheng, C.~Zheng, and X.~Li, ``{ICASSP 2021 Acoustic Echo
  Cancellation Challenge: Integrated Adaptive Echo Cancellation with Time
  Alignment and Deep Learning-Based Residual Echo Plus Noise Suppression},'' in
  \emph{Proceedings of ICASSP}.\hskip 1em plus 0.5em minus 0.4em\relax IEEE,
  2021.

\bibitem{watcharasupat2021end}
K.~N. Watcharasupat, T.~N.~T. Nguyen, W.-S. Gan, S.~Zhao, and B.~Ma,
  ``{End-to-End Complex-Valued Multidilated Convolutional Neural Network for
  Joint Acoustic Echo Cancellation and Noise Suppression},'' in
  \emph{Proceedings of ICASSP}.\hskip 1em plus 0.5em minus 0.4em\relax IEEE,
  2022, pp. 656--660.

\bibitem{ma2021echofilter}
L.~Ma, S.~Yang, Y.~Gong, X.~Wang, and Z.~Wu, ``{Echofilter: End-to-end Neural
  Network for Acoustic Echo Cancellation},'' \emph{arXiv preprint
  arXiv:2105.14666}, 2021.

\bibitem{ianniello1982time}
J.~Ianniello, ``{Time Delay Estimation via Cross-Correlation in the Presence of
  Large Estimation Errors},'' \emph{IEEE Transactions on Acoustics, Speech, and
  Signal Processing}, vol.~30, no.~6, pp. 998--1003, 1982.

\bibitem{clevert2015fast}
D.-A. Clevert, T.~Unterthiner, and S.~Hochreiter, ``{Fast and Accurate Deep
  Network Learning by Exponential Linear Units (ELUs)},'' in \emph{Proceedings
  of ICLR}, 2016.

\bibitem{peng2021acoustic}
R.~Peng, L.~Cheng, C.~Zheng, and X.~Li, ``{Acoustic Echo Cancellation Using
  Deep Complex Neural Network with Nonlinear Magnitude Compression and Phase
  Information},'' in \emph{Proceedings of INTERSPEECH}, 2021, pp. 4768--4772.

\bibitem{wisdom2019differentiable}
S.~Wisdom, J.~R. Hershey, K.~Wilson, J.~Thorpe, M.~Chinen, B.~Patton, and R.~A.
  Saurous, ``{Differentiable Consistency Constraints for Improved Deep Speech
  Enhancement},'' in \emph{Proceedings of ICASSP}.\hskip 1em plus 0.5em minus
  0.4em\relax IEEE, 2019, pp. 900--904.

\bibitem{ephrat2018looking}
A.~Ephrat, I.~Mosseri, O.~Lang, T.~Dekel, K.~Wilson, A.~Hassidim, W.~T.
  Freeman, and M.~Rubinstein, ``{Looking to Listen at the Cocktail Party: a
  Speaker-Independent Audio-Visual Model for Speech Separation},'' \emph{ACM
  Transactions on Graphics}, vol.~37, no.~4, pp. 1--11, 2018.

\bibitem{purin2021aecmos}
M.~Purin, S.~Sootla, M.~Sponza, A.~Saabas, and R.~Cutler, ``{AECMOS: A Speech
  Quality Assessment Metric for Echo Impairment},'' in \emph{Proceedings of
  ICASSP}.\hskip 1em plus 0.5em minus 0.4em\relax IEEE, 2022, pp. 901--905.

\bibitem{cutler2021crowdsourcing}
R.~Cutler, B.~Naderi, M.~Loide, S.~Sootla, and A.~Saabas, ``{Crowdsourcing
  Approach for Subjective Evaluation of Echo Impairment},'' in
  \emph{Proceedings of ICASSP}.\hskip 1em plus 0.5em minus 0.4em\relax IEEE,
  2021, pp. 406--410.

\bibitem{kingma2015adam}
D.~P. Kingma and J.~Ba, ``{Adam: A Method for Stochastic Optimization},'' in
  \emph{Proceedings of ICLR}, 2015.

\end{thebibliography}

\end{document}